\pdfoutput=1
\documentclass[aps,prl,twocolumn,showpacs,superscriptaddress]{revtex4-1}

\newcommand{\half}{\frac{1}{2}}

\usepackage[pdftex]{graphicx}
\usepackage[pdftex]{epsfig}
\usepackage{epstopdf}
\usepackage{color}
\usepackage{amssymb,amsmath}
\usepackage{graphicx}
\usepackage{dcolumn}
\usepackage{multirow}
\usepackage{hyperref}

\begin{document}

\def\simlt{\mathrel{\lower .3ex \rlap{$\sim$}\raise .5ex \hbox{$<$}}}
\def\half{ \frac {1}{2} }
\title{Valley splitting of single-electron Si MOS quantum dots}
  
\author{John King Gamble}
\email{jkgambl@sandia.gov}
\affiliation{Center for Computing Research, Sandia National Laboratories, Albuquerque, NM 87185, USA}
\author{Patrick Harvey-Collard}
\affiliation{D\'epartement de physique, Universit\'e de Sherbrooke, Sherbrooke, QC, J1K 2R1, Canada}
\affiliation{Institut quantique, Universit\'e de Sherbrooke, Sherbrooke, QC, J1K 2R1, Canada}
\affiliation{Sandia National Laboratories, Albuquerque, NM 87185, USA}
\author{N. Tobias Jacobson}
\affiliation{Center for Computing Research, Sandia National Laboratories, Albuquerque, NM 87185, USA}
\author{Andrew D. Baczewski}
\affiliation{Center for Computing Research, Sandia National Laboratories, Albuquerque, NM 87185, USA}
\author{Erik Nielsen}
\affiliation{Sandia National Laboratories, Albuquerque, NM 87185, USA}
\author{Leon Maurer}
\affiliation{Department of Physics, University of Wisconsin-Madison, Madison, WI 53706, USA}
\author{In\`es Monta\~no}
\affiliation{Sandia National Laboratories, Albuquerque, NM 87185, USA}
\author{Martin Rudolph}
\affiliation{Sandia National Laboratories, Albuquerque, NM 87185, USA}
\author{M. S. Carroll}
\affiliation{Sandia National Laboratories, Albuquerque, NM 87185, USA}
\author{C. H. Yang}
\affiliation{Australian Research Council Centre of Excellence for Quantum Computation and Communication Technology, School of Electrical Engineering \& Telecommunications, The University of New South Wales, Sydney 2052, Australia}
\author{A. Rossi}
\affiliation{Cavendish Laboratory, University of Cambridge, Cambridge, CB3 0HE, UK}
\author{A. S. Dzurak}
\affiliation{Australian Research Council Centre of Excellence for Quantum Computation and Communication Technology, School of Electrical Engineering \& Telecommunications, The University of New South Wales, Sydney 2052, Australia}
\author{Richard P. Muller}
\affiliation{Center for Computing Research, Sandia National Laboratories, Albuquerque, NM 87185, USA}

\begin{abstract}
Silicon-based metal-oxide-semiconductor quantum dots are prominent candidates for high-fidelity, manufacturable qubits. 
Due to silicon's band structure, additional low-energy states persist in these devices, presenting both challenges and opportunities.
Although the physics governing these valley states has been the subject of intense study, quantitative agreement between experiment and theory remains elusive.
Here, we present data from a new experiment probing the valley states of quantum dot devices and develop a theory that is in quantitative agreement with both
the new experiment and a recently reported one.
Through sampling millions of realistic cases of interface roughness, our method provides evidence that, despite radically different processing, the valley physics between the two samples 
is essentially the same.
This work provides the first evidence that valley splitting can be deterministically predicted and controlled in metal oxide semiconductor quantum dots, a critical requirement for such systems to realize a reliable qubit platform.
\end{abstract}

\maketitle

Qubits based on isolated electron spins in semiconductors are one of the earliest proposals for a quantum information processing  architecture \cite{Loss:1998},
where electrons are confined to zero-dimensional quantum dots via electrostatic gates patterned on the surface of a semiconductor heterostructure \cite{hanson:2007}.
These isolated electrons resemble artificial atoms, and are very versatile: recent work have extended the simple single-spin encoding to two- \cite{levy:2002,petta:2005}, three- \cite{divincenzo:2000,shi:2012}, and even a proposed four-spin encoding \cite{friesen:2016}.

Two promising material choices for these devices are GaAs \cite{petta:2005,koppens:2006,shulman:2012} or Si \cite{maune:2012,pla:2012,kim:2014b,veldhorst:2015}. 
GaAs-based heterostructures have higher mobility than Si-based devices, which, combined with a smaller effective mass, leads to easier fabrication and greater device reliability.
However, GaAs devices have modest intrinsic coherence properties as compared with Si, due to non-zero nuclear spin; in Si, electron spin coherence can easily range to seconds \cite{zwanenburg:2013}.

An additional complication in Si arises from its band structure;
in the bulk, the conduction band has six degenerate minima, called \emph{valleys}. 
This valley degeneracy is broken by the sharp material interfaces present in heterostructures, 
resulting in a low-lying manifold of additional electronic states.
The presence of these states can be either a benefit \cite{smelyanskiy:2005,culcer:2012} or a drawback \cite{eriksson:2004,goswami:2007,culcer:2009}, but understanding 
and being able to predictably engineer the valley physics in quantum dots is critically important for developing functional qubits.

For these reasons, the valley physics of silicon has been the subject of intense study over the past decade.
Researchers have used effective mass theory \cite{friesen:2007}, atomistic pseudopotentials \cite{zhang:2013}, and atomistic tight binding \cite{boykin:2004,rahman:2011a} to make predictions
of the energy gap between the lowest two valley states, termed the valley splitting, in a variety of experimentally-relevant scenarios.
These studies indicate that disorder in the heterostructure interface dramatically influences the valley splitting \cite{shi:2011,culcer:2010},
leading to the unfortunate conclusion that valley splitting may vary substantially amongst nominally identical devices.

Recently, experiments have advanced to the point where it is possible to track valley splitting as a function of applied electrostatic biases while maintaining single-electron dot occupation \cite{yang:2013}.
In this work, we present a second measurement on a device with a significantly different design and fabrication process.
We then develop a non-perturbative, multi-valley effective mass theory that can directly simulate both experiments.
We find similar, predictable behavior in the tuning of the valley splitting.
Our theory enables efficient high-throughput numerical sampling of random interfaces, achieving quantitative agreement with experiment and providing a substantial improvement upon previous work.

\begin{figure*}[tb]
\includegraphics[width= 1.0 \linewidth]{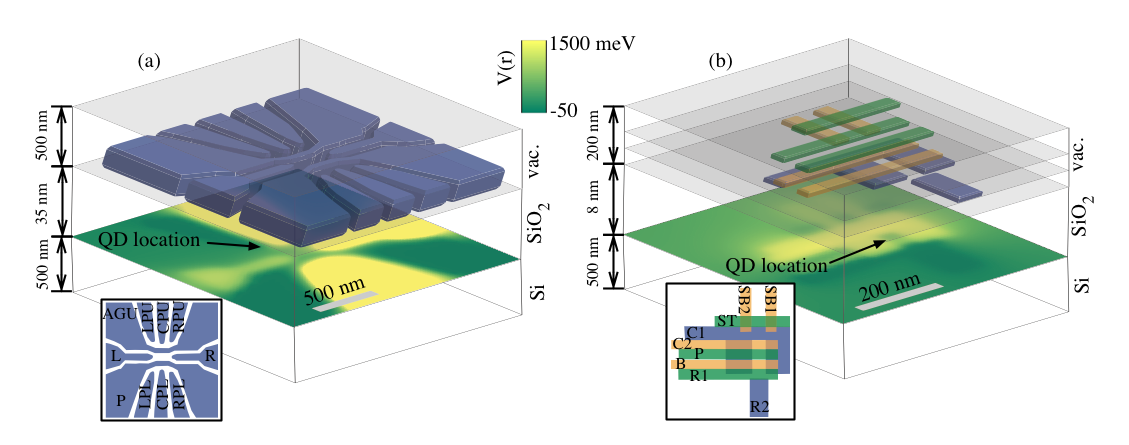}
\caption{\label{devSchematics}(Color online) 
Schematics of both devices used in this work.
(a) The device fabricated at Sandia National Laboratories, detailed in Ref.~\onlinecite{harvey-collard:2015}. 
This device consists of a single layer of 200 nm n-doped poly-silicon electrodes deposited on 35 nm oxide. 
The single quantum dot studied here is formed at the Si-oxide interface as indicated. 
(b) The device fabricated at the University of New South Wales, detailed in Ref.~\onlinecite{yang:2013}. 
This device uses three layers of aluminum electrodes, which are deposited on 8 nm oxide. The gates are separated by a thin layer of thermally grown Al$_\text{x}$O$_\text{y}$. In both cases, the potential energy surface $V(r)$ witnessed by electrons 3 nm beneath the oxide is shown, the result of self-consistent Thomas-Fermi calculations detailed in the main text.
}
\end{figure*}

The experiments were performed on two different samples, each a metal-oxide-semiconductor (MOS) quantum dot (QD) nanostructure. 
In both samples, electrodes patterned on the top of the device were used to provide electron confinement, isolating a single electron in a QD.
The first device, depicted in Fig.~\ref{devSchematics}(a), is a single-layer gated wire geometry fabricated at Sandia National Laboratories (SNL)\cite{harvey-collard:2015}. 
The second device, shown in Fig.~\ref{devSchematics}(b), is a three-layer design fabricated at the University of New South Wales (UNSW) \cite{yang:2013}.

In both experiments, the valley splitting of a single-electron quantum dot was measured. 
In tightly confined quantum dots like those considered here, the first excited state carries a valley-like degree of freedom \cite{gamble:2013}, 
so the valley splitting is given by the difference between the first excited and ground state of the quantum dot: $E_{VS} = E_1 - E_0$. 
Here, confinement to a narrow sheet next to the interface splits the six-fold degenerate conduction band minima of bulk silicon into a low-lying doublet and an excited quadruplet.
The doublet, whose conduction band minima lie along the $\pm \hat z$ directions in momentum space, is further split by the sharp oxide interface potential.
By changing the voltages on the control electrodes while compensating to ensure the quantum dot remains in the single-electron regime,
the electronic wavefunction can be forced to penetrate more into the oxide barrier, effectively tuning the valley splitting as a function of voltage configuration.

In the SNL experiment, the valley splitting is measured using a pulsed gate spectroscopy technique \cite{nakamura:1999,petta:2005}. 
The quantum dot is tuned to the single electron regime and the tunnel rate to the lead is adjusted to roughly 10 kHz. 
Then, using a square pulse on the CPL gate (Fig.~\ref{devSchematics}) of varying frequency and mean voltage, the excited one-electron states of the quantum dot 
are probed by monitoring the average quantum dot occupancy with the charge sensor \cite{shi:2011}. For a small range of frequencies,
 both the ground state and the first excited state can be seen, even if they have similar tunnel rates, allowing the measurement of their energy separation.
  Gate voltage differences are converted to energy using a gate lever-arm which is calibrated through a temperature dependence measurement.
  The procedure is repeated for multiple gate voltage configurations. 
  In particular, the P gate has the most influence on the electrical field perpendicular to the oxide interface, and thus has a large influence on the valley splitting as well. 
  The gate voltages were thus chosen to explore a large range of P voltages.

In the UNSW experiment, reported in Ref.~\onlinecite{yang:2013},
two techniques were used to measure the valley splitting. For small plunger gate voltages $V_p$, a spin-relaxation hotspot, for which the electron $T_1$ time is minimized when the valley splitting is commensurate with the Zeeman splitting of the device, was used. For  larger values of $V_p$, pulsed-gate magnetospectroscopy was used. For details  regarding the measurement technique, see Ref.~\onlinecite{yang:2013}.

Since the valley splitting depends sensitively on the electrostatics of the problem, here we take a multi-stage approach to our calculations. 
First, we perform self-consistent Thomas-Fermi simulations \cite{stopa:1996} of the devices under the experimental voltage configurations.
In the leads of the device, this simulation captures the effect of dynamic screening, using COMSOL Multiphysics with a 2D density of states to self-consistently model charge accumulation at the oxide-silicon interface
Nearby the quantum dots, we exclude self-consistent accumulation, since the devices are experimentally tuned to the single-electron regime.
\footnote{
When tuned using the experimental voltages, the UNSW device simulation exhibited a clear quantum dot potential, so the dot accumulation reason was readily excluded from accumulation.
The SNL simulation did not show a clear dot-lead separation, so the choice of exclusion zone (which then generates a dot confinement potential), was somewhat ambiguous. A possible explanation for this discrepancy is non-uniform fixed charge in the real device but not in the simulation.}.

The output of the electrostatic simulations is then fed into a non-perturbative multi-valley effective mass theory \cite{gamble:2015}. Within this framework, the electronic wave function is assumed to have momentum-space support only nearby the two low-lying conduction band minima:
\begin{equation}
\psi(\mathbf{r}) = \sum_{j=1}^2 F_j(\mathbf r) \phi_j(\mathbf r),
\end{equation}
where $\mathbf k_0^j$ indexes the valley minima (located $0.84 \times 2 \pi/a$ along the $z$ axes, with $a=0.543$~nm the cubic unit cell length of Si), $F_j$ is the envelope function of the $j$th valley, and $\phi_j$ is the Bloch function of the $j^{th}$ valley (see Supplemental Information).
The wave function satisfies a system of coupled Schr\"odinger equations \cite{shindo:1976},
\begin{equation}\label{eq:SN}
E  F_l(\mathbf r)  = \left( \hat{\mathbf{T}}_l + U(\mathbf r ) \right) F_l(\mathbf r) + \sum_{j =1}^2 V^{VO}_{lj}(\mathbf r) F_j(\mathbf r),
\end{equation}
where $E$ is the energy, $\hat{\mathbf{T}}_l$ is the kinetic energy operator of the $l^{th}$ valley, $U(\mathbf r)$ is the potential energy landscape for the electron,
and $V^{VO}_{lj}(\mathbf{r}) = \phi_l^*(\mathbf r) \phi_j(\mathbf r)  U(\mathbf r)$ is the valley-orbit coupling.

To solve Eq.~(\ref{eq:SN}), we expand the envelopes $F_l$ in a fixed orbital basis set,
\begin{equation}
 F_j(\mathbf r) = \sum_{x_0,y_0,z_0} A_{(j,x_0,y_0,z_0)} G_{(j,x_0,y_0,z_0)}(\mathbf r),
\end{equation}
where the indices $(x_0,y_0,z_0)$ index a real-space grid Gaussian basis (see Supplemental Information).

\begin{figure}[tb]
\includegraphics[width= 1.0 \linewidth]{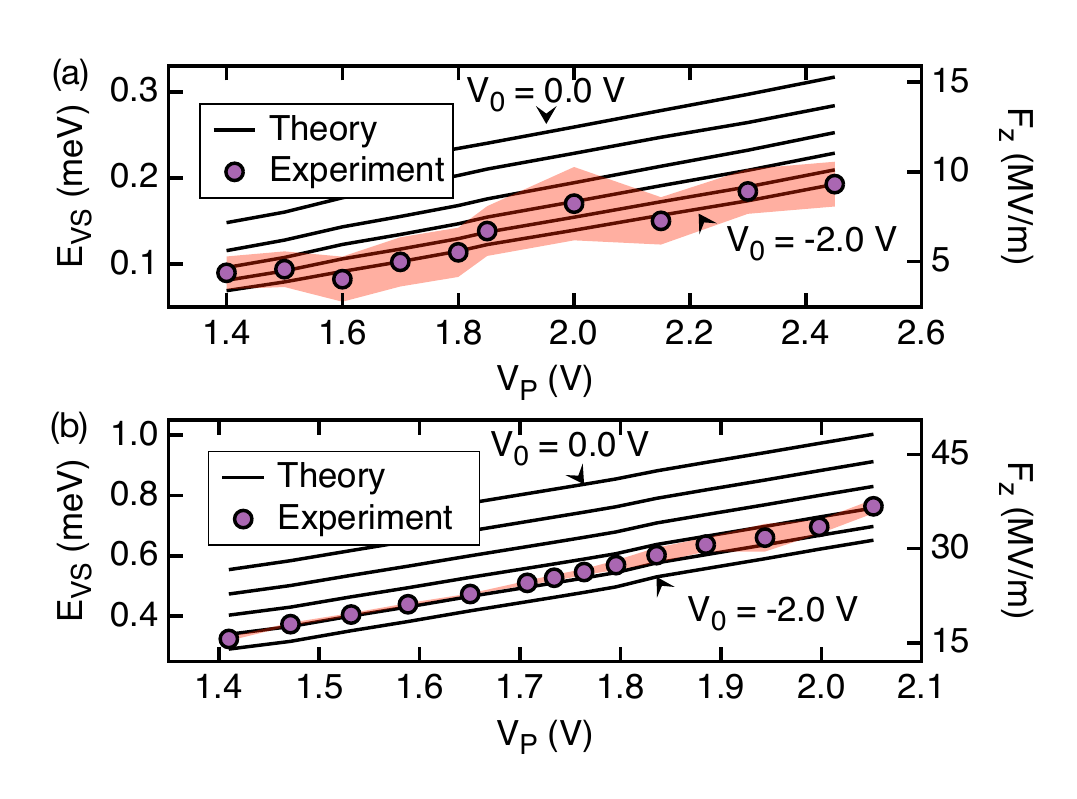}
\caption{\label{valleySplittingCalcs}(Color online) 
Measured and computed valley splittings using multi-valley effective mass theory and an ideal, sharp interface.
(a) Results for the SNL device \cite{harvey-collard:2015}. The experimental data (dots) are shown with measurement error bars as a band. 
The theory curves are computed using the experimental voltages and an additional uniform voltage offset $V_0$, generating a family of curves evenly distributed between 0 V and -2 V. 
(b) Results from the UNSW device \cite{yang:2013}. 
The experimental data and error bars were reported in Ref.~\onlinecite{yang:2013}; the theory results are new here.
In both cases, the best-fit voltage offset is about $V_0 = -1.8$ V, which is significantly more than typical threshold voltage shifts observed in experiment.
A possible explanation of this discrepancy is interface disorder, explored in Fig.~\ref{interfaceDisorder}.
The effective vertical electric fields plotted on the right axis are computed using the vertical field vs. valley splitting relationship established in Fig.1(b) of the Supplemental Information.
}
\end{figure}
\begin{figure*}[tb]
\includegraphics[width= 1.0 \linewidth]{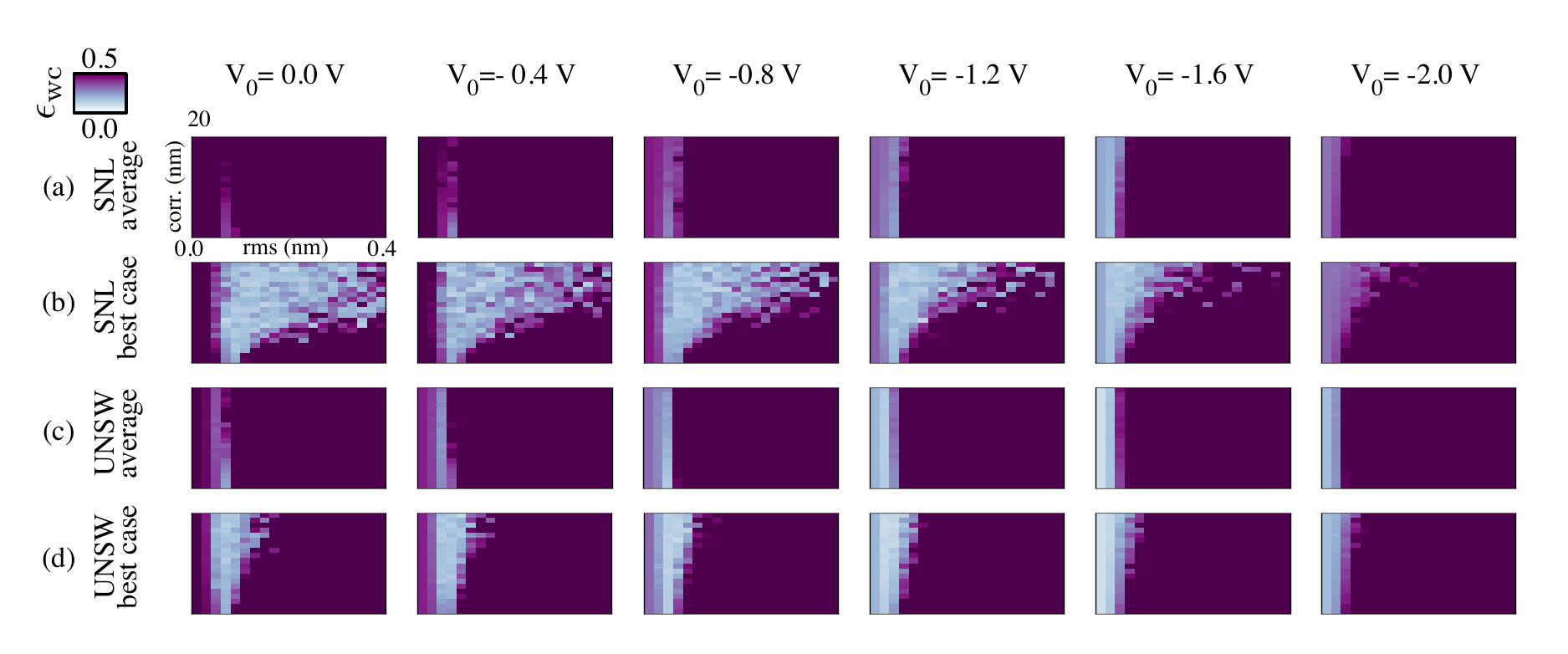}
\caption{\label{interfaceDisorder}(Color online) 
Worst case relative error $\epsilon_\text{wc}$ of valley splitting with respect to experiment for disordered interfaces.
Each plot shows a 2D sweep over correlation length and root-mean-squared (RMS) roughness, with each parameter pair having 65 samples.
Offset voltages between $V_0 = 0.0$~V and $V_0 = -2.0$~V are shown as columns, which correspond to the lines shown in Fig.~\ref{valleySplittingCalcs}.
The values reported correspond to the worst case relative error, taken over all the data points of the experiment.
We report  $\epsilon_\text{wc}$ averaged over the interface realizations in panels (a) and (c) and for the best interfaces in panels (b) and (d). 
By including disorder, we see that our theory is consistent with the smaller threshold voltages see in experiment.
}
\end{figure*}

We simulated both experiments using multi-valley effective mass theory, as described above. 
First, we restricted ourselves to the case of a flat interface. 
Our electrostatic simulations do not take the threshold voltage shift due to oxide charge into account directly.
To mimic various thresholds, we apply a uniform voltage shift $V_0$ to all the electrodes in the simulation.

In Fig.~\ref{valleySplittingCalcs}, we show the results of these calculations.
For both devices, the computed slope of valley splitting vs. voltage agrees well with experiment.
For no voltage offset ($V_0 = 0$), there is a pronounced uniform shift of ~0.1-0.2 meV, with the theory overestimating
the valley splitting.
This is not unexpected: an offset was previously observed \cite{yang:2013},
where it was attributed to interface disorder.
In that previous work, the offset was reported to be considerably larger than what we find here (~1 meV). 
Our model directly computes the valley splitting from the full electrostatic potential, including the important fringing, non-uniform vertical field,
directly computing the valley splitting from the electrostatic potential.
In contrast, Ref.~\onlinecite{yang:2013} extracted an approximate vertical electric field from TCAD calculations and 
fed the results into previous simple model system calculations of valley splittings.
To make more direct contact with previous results, we can translate our valley splitting results
into an effective vertical electric field (\emph{i.e.}, the vertical electric field that, in an ideal model system, would explain the valley splitting). 
We do this using the valley splitting vs. vertical electric field results shown in Fig.~1(b) of the Supplemental Information,
and we show the effective vertical electric fields in Fig.~\ref{valleySplittingCalcs} on the right axes. 
The UNSW device exhibits higher valley splittings than the SNL device mainly due to thinner oxide thickness, smaller device features, and larger applied voltages,
all of which serve to raise the effective vertical electric field.

Despite obtaining excellent experimental agreement at $V_0 = -1.8$~V, thresholds in these devices are typically between 0.1-1.0~V.
Hence, from experiment we expect to need to include a compensating offset of $V_0 = -0.1$ to $-1.0$~V in our simulations.
To investigate this apparent discrepancy, in Fig.~\ref{interfaceDisorder} we show the effect of disordered interfaces on the valley splitting. 
We parameterize the interface using a Gaussian correlation function and a two-parameter correlation length and RMS roughness model \cite{culcer:2010}.
We sample these parameters over a 20x20 grid, with 65 random realizations per point.
For each case, we choose a voltage offset $V_0$ and then compute the valley splitting for the experimental voltages.
We report the worst-case relative error $\epsilon_\text{wc}$ with respect to the experimental valley splittings, defined as:
\begin{equation}
\epsilon_\text{wc} = \max_{V_p} \frac{\left| E_{VS}^\text{exp}(V_p) - E_{VS}^\text{theory}(V_p) \right|}{ \left| E_{VS}^\text{exp}(V_p) \right|},
\end{equation}
where $E_{VS}^\text{exp}(V_p)$ and $E_{VS}^\text{theory}(V_p)$ are the measured and predicted valley splittings at voltage $V_p$, respectively.

For both the SNL and UNSW devices, we show $\epsilon_\text{wc}$ averaged over the 65 interface realizations as well as the result for the best interface. 
In both cases, we found disordered interfaces that are consistent with
the lower threshold voltages observed in experiment as well as realistic MOS interface parameters of RMS roughness $\sim0.1$~nm \cite{goodnick:1985} and a wide range of correlation lengths.
This shows that the introduction of realistic disorder is sufficient to solve the apparent discrepancy between theoretical and experimental threshold voltages noted in Fig.~\ref{valleySplittingCalcs}.

In this work, we analyzed the valley splitting for two distinct MOS devices: a single-layer gated-wire design fabricated at SNL,
and a multi-layer device fabricated at UNSW. 
Despite superficially appearing to have very different valley splitting properties, detailed MVEMT calculations of the valley splitting, directly incorporating
the potential energy landscape, revealed that geometric differences are likely responsible for the differences and that the valley physics is consistent across the two devices.
By introducing a voltage offset of -1.8~V to mimic threshold voltage, we obtained quantitative agreement with experiment.
Since this value is larger than what is typically seen in experiment, we implemented a non-perturbative disordered interface model to attempt to explain this discrepancy.
Through this, we found plausible interface roughness parameters that lead to realistic threshold voltages.

Overall, our results suggest that MOS quantum dots are a promising qubit platform.
Since excessively small valley splitting is problematic for qubit operation, being able to reliably tune and design for large valley splitting is critical for successful qubit operation.
Here, we have put forward evidence that MOS single-electron valley splitting is both \emph{tunable} and \emph{predictable}, opening the door to further design and optimization 
of robust qubits.

The authors acknowledge useful discussions with F. Mohiyaddin and M. Usman. 
Sandia National Laboratories is a multi-program laboratory managed and operated by Sandia Corporation, a wholly owned subsidiary of Lockheed Martin Corporation, for the U.S. Department of Energy's National Nuclear Security Administration under contract DE-AC04-94AL85000.  
JKG gratefully acknowledges support from the Sandia National Laboratories Truman Fellowship Program, which is funded by the Laboratory Directed Research and Development (LDRD) program. 
This work was performed, in part, at the Center for Integrated Nanotechnologies, an Office of Science User Facility operated for the U.S. Department of Energy (DOE) Office of Science.
C.H.Y. and A.S.D.  acknowledge support from the Australian Research Council (CE11E0001017), the US Army Research Office (W911NF-13-1-0024) and the NSW Node of the Australian National Fabrication Facility.
AR acknowledges support from the European Union's Horizon 2020 research and innovation
programme under the Marie Sk\l{}odowska-Curie grant agreement No 654712 (SINHOPSI).

\appendix
\section{Supplemental Information}

\subsection{Bloch function and convergence details}

When computing the matrix elements required to solve Eq.~\ref{eq:SN}, we need to evaluate the Bloch functions of silicon, which can be written as a Fourier decomposition
\begin{equation}
\phi_j(\mathbf r) = \sum_{\mathbf G} A_{\mathbf G}^j e^{i (\mathbf k_0^j \mathbf G) \cdot \mathbf r},
\end{equation}
where $\mathbf G$ are reciprocal lattice vectors, $\mathbf k_0^j$ is a conduction band minimum location, and the coefficients $A_{\mathbf G}^j$ are determined from density functional
theory \cite{saraiva:2011,gamble:2015}.
In particular, we need to evaluate pair products of Bloch functions, which have the form
\begin{equation}
\phi_l^*(\mathbf r) \phi_j(\mathbf r)  =  \sum_{\mathbf G, \mathbf G'} \left(A_{\mathbf G'}^l \right)^* A_{\mathbf G}^j e^{i (\mathbf k_0^j-\mathbf k_0^l+\mathbf G - \mathbf G' ) \cdot \mathbf r}.
\end{equation}
Since the difference of reciprocal lattice vectors is itself a reciprocal lattice vector, we can rewrite this as
\begin{equation}
\phi_l^*(\mathbf r) \phi_j(\mathbf r)  =  \sum_{\mathbf{\Delta G}} \alpha_{\mathbf{\Delta G}}^{j,l} e^{i (\mathbf k_0^j-\mathbf k_0^l+\mathbf{\Delta G} ) \cdot \mathbf r},
\end{equation}
where the coefficients $\alpha_{\mathbf{\Delta G}}^{j,l}$ are determined from the appropriate combinations of products of $A_{\mathbf G}^j$ coefficients.

\begin{figure}[tb]
\includegraphics[width= 1.0 \linewidth]{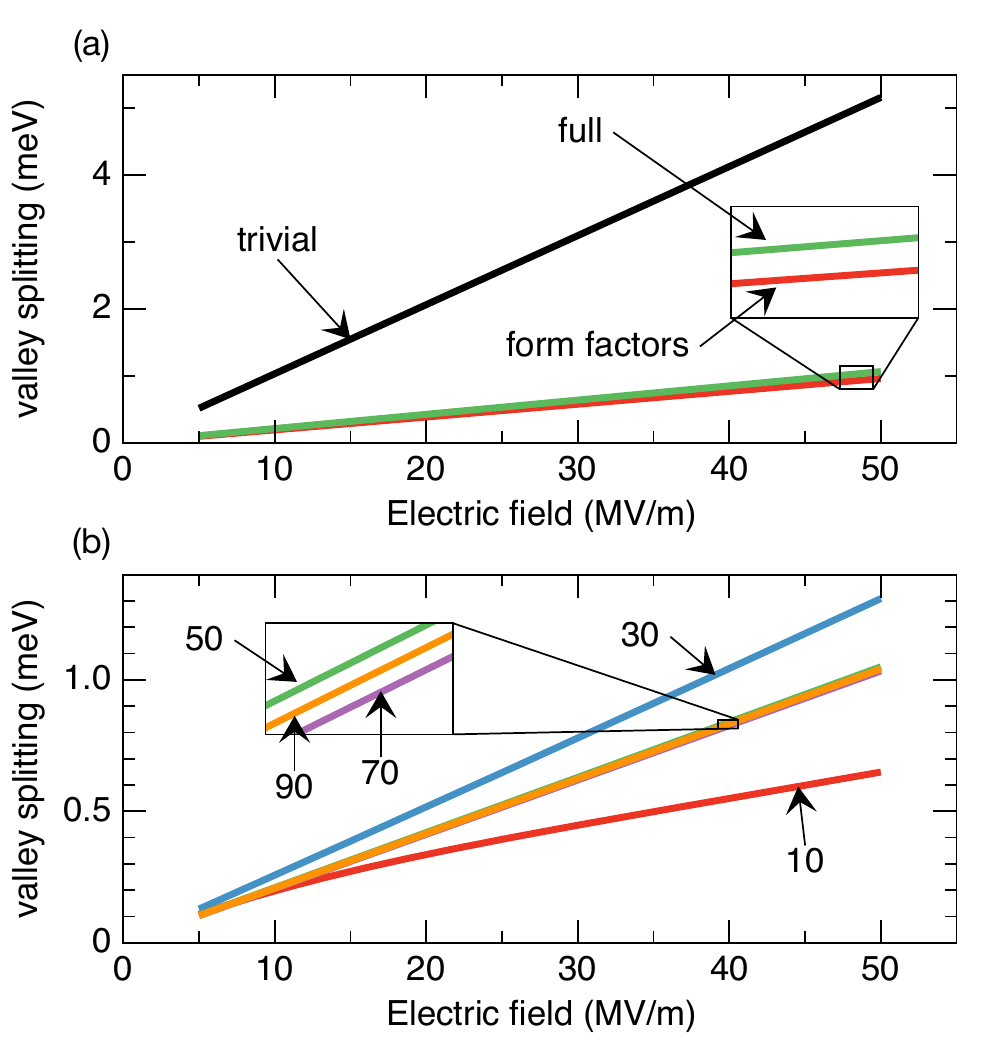}
\caption{\label{convTest}(Color online) 
Convergence tests for the valley splitting calculations.
(a): Valley splitting as a function of vertical electric field for three different approximations of the Bloch functions. 
In contrast to the case for donors in silicon, the difference between form factors and full Bloch Functions is quite small.
(b): Valley splitting as a function of vertical electric field for different numbers of Gaussian basis functions in the $z$ direction. 
By about 50 Gaussians, the calculation has converged.
}
\end{figure}

Since in practice only a finite number of $\mathbf{\Delta G}$ terms may be included, we need to truncate the series after some threshold. In previous work \cite{gamble:2015}, 
it was necessary to retain all terms out to $\Delta G \leq 4.4 \times 2 \pi / a$. 
However, that work considered a phosphorus donor, which is quite different from a quantum dot. 
Since including a large number of terms dramatically increases computational cost, in Fig.~\ref{convTest}~(a) we perform an analysis comparing three different approximation schemes for Bloch functions:
the trivial case where $\phi_j(\mathbf r) = e^{i \mathbf k_0^j  \cdot \mathbf r}$, the form factor case where only the $\alpha_{\mathbf{ 0}}^{j,l}$ term is retained, and the full case where 
all  $\Delta G \leq 4.4 \times 2 \pi / a$ are kept. 
For simplicity, here we considered only a model quantum dot with a linear applied electric field.
This test shows that using the form factor approximation is adequate to capture the physics of the valley splitting of a quantum dot at an interface.

Next, we performed convergence testing to calibrate the number of $z$-Gaussians necessary to resolve the wavefunction finely enough in the vicinity of the interface to converge the valley splitting. 
For the purposes of this test, we space the Gaussians unifromly between $z=-13$~nm and $z=+3$~nm, with the interface at $z=0$.
In Fig.~\ref{convTest}~(b) we show results for $n=10,30,50,70,90$ Gaussians.
We see convergence after about $n=50$, indicating that $n=60$, as used in the main text, is adequate for good convergence.

\subsection{Bloch function and convergence details}
Each individual basis function has the form
\begin{equation}\label{eq:gridbasis}
G_{(j,x_0,y_0,z_0)}(\mathbf r) = e^{-\frac{(x-x_0)^2}{(2 L_x^2)}}
e^{-\frac{(y-y_0)^2}{(2 L_y^2)}}
e^{-\frac{(z-z_0)^2}{(2 L_z^2)}},
\end{equation}
where the widths $L_x$,$L_y$, and $L_z$ are chosen to be 80\% of the grid spacing.
We use a total of 7 grid points along $x$ and $y$, and 60 grid points along $z$, 
with a computational domain that is $300\times175\times10$ nm for the SNL device and 
$40\times36\times10$ nm for the UNSW device. 
Hence, the total basis size is 2940 Gaussians per valley, with both $z$ valleys considered.
In both cases, the Gaussian grid is set back from the simulation domain wall such that $G$ evaluated at the boundary is at most $1\times 10^{-6}$.
Note that the grid basis is much finer and more dense along the $z$ axis, since resolving the shape along this axis is critical to accurately resolve the valley splitting.

The potential energy in our problem is
\begin{equation}
U(\mathbf r) = U_{BG}(\mathbf r) + U_\text{wall},
\end{equation}
where $U_{BG}$ is the background electrostatic potential, obtained from COMSOL simulations, and $U_\text{wall}$ is the interface potential, which is taken to be an abrupt offset of 3 eV. 
Interface roughness is incorporated into the simulations directly by considering a $200\times 200$ ``pixelated" array of wall positions in the $x-y$ plane, the positions of which were determined 
randomly according to a Gaussian correlation function, for various correlation and root-mean-square (RMS) parameters.
We then posed the entire problem, including the disorder matrix elements, as a generalized eigenvalue problem of dimension 5880 and diagonalized, solving for the low-lying states using an iterative eigenvalue procedure. 

\subsection{Bloch function and convergence details}
Since we aim to solve the eigenvalue problem non-perturbatively using a large, fixed basis, we need to efficiently compute the matrix elements of the interface with all pairs of basis functions. 
First, the wall potential is given by
\begin{equation}
U_\text{wall}(\mathbf R) = U_0 \Theta(z - \zeta(\mathbf r)),
\end{equation}
where $\zeta(\mathbf r)$ gives the z position of the interface as a function of lateral coordinates.
Since our grid basis defined in Eq.~\ref{eq:gridbasis} is separable along the cartesian directions, it is convenient to decompose it into planar and $z$ components:
\begin{equation}
F_{ij}(\mathbf R) = f_i(\mathbf r) g_j(z),
\end{equation}
where there are $m$ such $f_i$ radial orbitals and there are $n$ $g_j$ z orbitals. Because we are working with a regular grid, thre are $m\times n$ $F_{ij}$ 3D orbitals. Our goal is to compute the matrix elements
\begin{align}
A(i',j',k';i,j,k) = \left< F_{i'j'},k' \right| U_\text{wall} \left| F_{ij},k \right>,
\end{align}
where $k$ and $k'$ are valley indices. Unfortunately, these matrix elements are \emph{not} immediately separable, so we need to do some work to compute them. We have
\begin{widetext}
\begin{align}
A(i',j',k';i,j,k) &= \left< F_{i'j'},k' \right| U \left| F_{ij},k \right> \\
&= \sum_{\mathbf G,\mathbf G'}c_k(\mathbf G)c_{k'}^*(\mathbf G') 
\int d^3 R e^{i\mathbf R \cdot (\mathbf G + \mathbf k_0^k - \mathbf G' - \mathbf k_0^{k'})}
F_{ij}(\mathbf R) F_{i'j'}(\mathbf R)U_0 \Theta(z - \zeta(\mathbf r)) \nonumber\\
&=\sum_{\mathbf{\Delta G}} \alpha_{(\mathbf{ \Delta G_r},\Delta G_z)}^{k,k'}
\int d^3 R e^{i\mathbf R \cdot (\mathbf{\Delta G} + \mathbf{\Delta k})}
F_{ij}(\mathbf R) F_{i'j'}(\mathbf R)U_0 \Theta(z - \zeta(\mathbf r))\nonumber\\
&= U_0 \sum_{\mathbf{\Delta G_r}, \Delta G_z} \alpha_{(\mathbf{ \Delta G_r},\Delta G_z)}^{k,k'}
\int d^2 r e^{i\mathbf r \cdot (\mathbf{\Delta G_r} + \mathbf{\Delta k_r})} f_i(\mathbf r) f_{i'}(\mathbf r)
\int dz e^{iz(\Delta G_z + \Delta k_z)}g_j(z) g_{j'}(z)\Theta(z - \zeta(\mathbf r)) \nonumber\\
&\equiv U_0 \sum_{\mathbf{\Delta G_r}, \Delta G_z} \alpha_{(\mathbf{ \Delta G_r},\Delta G_z)}^{k,k'}
\int d^2 r e^{i\mathbf r \cdot (\mathbf{\Delta G_r} + \mathbf{\Delta k_r})} f_i(\mathbf r) f_{i'}(\mathbf r)
\lambda\left(\Delta G_z + \Delta k_z,g_jg_{j'},\zeta(\mathbf r)\right)\nonumber\\
& = U_0 
\int d^2 r \left[ \sum_{\mathbf{\Delta G_r}, \Delta G_z} \alpha_{(\mathbf{ \Delta G_r},\Delta G_z)}^{k,k'}
 e^{i\mathbf r \cdot (\mathbf{\Delta G_r} + \mathbf{\Delta k_r})} f_i(\mathbf r) f_{i'}(\mathbf r)
\lambda \left(\Delta G_z + \Delta k_z,g_jg_{j'},\zeta(\mathbf r)\right) \right]\nonumber \\
&\approx  U_0 
\int d^2 r \left[  \alpha_{\mathbf{ 0}}^{k,k'}  f_i(\mathbf r) f_{i'}(\mathbf r)
\lambda \left( \Delta k_z,g_jg_{j'},\zeta(\mathbf r)\right) \right],\nonumber 
\end{align}
\end{widetext}
where we have taken the form factor approximation in the last step by only retaining the $\mathbf{\Delta G} = \mathbf 0$ term.
We note that here we consider only $\mathbf{\Delta k_r} = \mathbf 0$ cases, since the x and y valleys are energetically split off from the z valleys. 
Above, we defined the function
\begin{equation}
\lambda(k,g,z_i) \equiv \int_{z_i}^{\infty} dz e^{i k z}g(z),
\end{equation}
which we pre-compute over a grid of wall positions. 
Since our underlying basis is a collection of Gaussians, the 1D $\lambda$ integrals are carried out analytically.
We perform the remaining 2D integral numerically using the trapezoid rule. 

\bibliography{refs.bib}

\end{document}